\documentclass[two column,rsi,amsmath,achemso,superscriptaddress,pre]{revtex4-1}
\setcitestyle{super}
\usepackage{color,graphics}
\usepackage{graphicx}
\usepackage[sort&compress]{natbib}
\usepackage{hhline}
\usepackage{soul}
\usepackage{xcolor}
\usepackage{array}
\usepackage{bm}

\usepackage[normalem]{ulem}
\usepackage{hyperref}
\usepackage[thinlines]{easytable}

\begin{document}

\setcounter{totalnumber}{3}
\renewcommand{\thetable}{\arabic{table}}
\newcolumntype{P}[1]{>{\centering\arraybackslash}p{#1}}

\title{Nonlocal screening dictates the radiative lifetimes of excitations\\ in lead halide perovskites}

\author{Yoonjae Park}
 \affiliation{Department of Chemistry, University of California, Berkeley, California 94720, United States}

\author{Amael Obliger}
 \thanks{Current Address: ISM, Univ. Bordeaux, CNRS, Talence, France}
 \affiliation{Department of Chemistry, University of California, Berkeley, California 94720, United States}
 
\author{David T. Limmer}
 \email{dlimmer@berkeley.edu}
 \affiliation{Department of Chemistry, University of California, Berkeley, California 94720, United States}
\affiliation{Materials Science Division, Lawrence Berkeley National Laboratory, Berkeley, California 94720, United States}
\affiliation{Chemical Science Division, Lawrence Berkeley National Laboratory, Berkeley, California 94720, United States}
\affiliation{Kavli Energy NanoScience Institute, Berkeley, California 94720, United States}

\date{\today}
\vspace{0mm}

\begin{abstract}
We use path integral molecular dynamics simulations and theory to elucidate the interactions between charge carriers, as mediated by a lead halide perovskite lattice. We find that the charge-lattice coupling of MAPbI$_3$ results in a repulsive interaction between electrons and holes at intermediate distances. The effective interaction is understood using a Gaussian field theory, whereby the underlying soft, polar lattice contributes a nonlocal screening between quasiparticles. Path integral calculations of this nonlocal screening model are used to rationalize the small exciton binding energy and low radiative recombination rate observed experimentally and are compared to traditional Wannier-Mott and Fr{\"o}hlich models, which fail to do so. These results clarify the origin of the high power conversion efficiencies in lead halide perovskites. Emergent repulsive electron-hole interactions provide a design principle for optimizing soft, polar semiconductors.\\[5pt]
%
%
\end{abstract}

\maketitle


Lead halide perovskites are a class of materials that have unique photophysical properties resulting from their soft, polar lattices. 
They have vanishingly small exciton binding energies and despite modest mobilities, have large free carrier diffusion lengths resulting from exceptionally long carrier lifetimes.\cite{science.aaa5333, adma.201305172,stranks2013electron} These properties make lead halide perovskites ideal materials for photovoltaic devices.\cite{nmat3911,aenm.201502458,lee2012efficient}
Many of their optoelectronic properties have been thought to arise from electron-phonon coupling, as the largely ionic bonding of the lead halides admit strong Coulomb interactions between free charges and the lattice.\cite{acs.jpclett.6b01425} It has been conjectured that polaronic effects in particular\cite{Zhu:2015eb4, ncomms12253, jacs.1c02403} act to protect free charges from recombination and screen their interactions, reducing exciton binding energies.\cite{ncomms12253} However, the significant anharmonicity of the perovskite lattice has made uncovering the molecular origin of these properties difficult.\cite{10.1038nat,ncomms8026, physrevb.92.144308,acs.jpclett.7b02423}

Here, we apply path integral molecular dynamics\cite{chandler1981exploiting,ceperley1995path} to study an atomistic model of quasiparticles embedded in a MAPbI$_3$ lattice, in order to understand how a fluctuating lattice affects its electronic properties. 
Much recent effort has gone into understanding the effects of the lattice on the excitonic properties of perovskites computationally \cite{acs.jpclett.9b02491, multiphonon, c8ee03369b,mayers2018lattice} and analytically.\cite{pssr.201510265, multiphonon} 
However, unlike traditional polar semiconductors where lattice fluctuations can be described by a harmonic approximation, the tilting and rocking motions of the inorganic octahedra\cite{acs.jpclett.7b02423} and nearly free motions of the A-site cations\cite{ncomms8026, physrevb.92.144308} render the lattice highly anharmonic. This complicates the simplification to traditional model Hamiltonians like the Fr{\"o}hlich model or its generalizations.\cite{mayers2018lattice,egger2018remains} Attempts to include lattice effects into \emph{ab initio} based approaches have been developed but these are difficult to extend to the time and length scales necessary to explain the nature of how photogenerated electrons and holes bind, dissociate and recombine.\cite{neatonprl} 
Using an explicit atomistic representation of the lattice surrounding the quasiparticles allows us to go beyond simplified models. Employing path integral calculations allows us to consider finite temperature effects directly on diffusive time and length scales. These simulations motivate a field theory to describe the effective electron-hole interactions that dictate the emergent optical properties of the perovskites. With these simulations and theory, we are able to elucidate the origin of low exciton binding energies and recombination rates as a consequence of a nonlocal screening from the lattice.

We consider a system of an electron-hole pair, interacting with a MAPbI$_3$ perovskite {\color{black}in its cubic phase} employing a fully atomistic description of the lattice. The full system Hamiltonian, $\mathcal{H}$, consists of electronic, lattice, and interaction pieces, $\mathcal{H} = \mathcal{H}_{\mathrm{el}} + \mathcal{H}_{\mathrm{l}} + \mathcal{H}_{\mathrm{int}}$.
The highly dispersive bands of MAPbI$_3$ allow us to make an effective mass approximation, so that the electronic Hamiltonian is defined as 
\begin{equation}
\begin{aligned}
\mathcal{H}_{\mathrm{el}} = \frac{\hat{\mathbf{p}}_e^2}{2m_e} + \frac{\hat{\mathbf{p}}_h^2}{2m_h} - \frac{e^2}{4\pi \varepsilon_{0} \varepsilon_{\infty} |\hat{\mathbf{r}}_e - \hat{\mathbf{r}}_h|}
\end{aligned}
\end{equation}
where the subscripts $e$ and $h$ indicate electron and hole, $\hat {\mathbf{p}}$ and $\hat {\mathbf{r}}$ are the momentum and position operators, $m_e/m=m_h/m=0.2$ are the band masses of the quasiparticles taken from recent GW calculations in units of the bare electron mass $m$,\cite{cho2019optical} $\varepsilon_{0}$ is the vacuum permittivity while $\varepsilon_{\infty}$ is the optical dielectric constant for charge $e$. For the lattice, we use an atomistic model developed by Mattoni et al., that has been demonstrated to reproduce the structural and dielectric properties of lead halide perovskites.\cite{mattoni,mattoni2020dielectric} Its Hamiltonian is decomposable as 
\begin{equation}
\begin{aligned}
\mathcal{H}_{\mathrm{l}} = \sum_{i=1}^N \frac{\hat{\mathbf{p}}_i^2}{2m_i} + U_{\mathrm{l}}(\hat{\mathbf{r}}^N)
\end{aligned}
\end{equation}
where $\hat{\mathbf{p}}_i$, $\hat{\mathbf{r}}_i$ and $m_i$ are the momentum, position, and mass of $i^{th}$ atom, $N$ is the total number of atoms in the lattice, and $U_{\mathrm{l}}(\hat{\mathbf{r}}^N)$  is the pair-wise interaction potential between atoms with configuration $\hat{\mathbf{r}}^N = \{\hat{\mathbf{r}}_1, \hat{\mathbf{r}}_2, \dots, \hat{\mathbf{r}}_N \}$. The potential includes electrostatic and excluded volume interactions. The charge-lattice interaction term is given by 
$\mathcal{H}_{\mathrm{int}} = U_{e,\mathrm{l}}(\hat{\mathbf{r}}_e,\hat{\mathbf{r}}^{N}) + U_{h,\mathrm{l}}(\hat{\mathbf{r}}_h,\hat{\mathbf{r}}^N)$ where $U_{e,\mathrm{l}}$ and $U_{h,\mathrm{l}}$ denote sums of pseudopotentials. Consistent with the largely ionic nature of MAPbI$_3$, we employ pseudopotentials of the form of short-ranged truncated Coulomb potentials,  with a cut-off radii chosen as the ionic radii of each species.\cite{parrinello1984study,schnitker1987electron,kuharski1988molecular}

{\color{black}As the atoms are heavy and we are largely interested in room temperature behavior, we adopt a classical description of the MAPbI$_3$ lattice. We discuss below corrections to this classical approximation in the harmonic lattice limit. For the two light quasiparticles however, we employ a path integral description to account for quantum mechanical effects important even at room temperature. Such a quasiparticle path integral approach has been employed previously to study lattice effects} in the lead halides and trapping in other semiconductors.\cite{bischak2017origin,bischak2018tunable,limmer2020photoinduced,remsing2020effective} The partition function, $\mathcal{Z}$, for the composite system can be written as 
\begin{equation}
\begin{aligned}
\mathcal{Z} = \int \mathcal{D} [{\mathbf{r}}_e,{\mathbf{r}}_h,\mathbf{r}^{N}] \, e^{-\mathcal{S}[{\mathbf{r}}_e,{\mathbf{r}}_h,\mathbf{r}^{N}]/\hbar}
\end{aligned}
\end{equation}
with the action $\mathcal{S}[{\mathbf{r}}_{e},{\mathbf{r}}_h,\mathbf{r}^{N}] = \mathcal{S}_{\mathrm{el}} + \mathcal{S}_{\mathrm{l}} + \mathcal{S}_{\mathrm{int}}$.
The corresponding imaginary time path action for the electronic part becomes
\begin{equation}
\begin{aligned}
\mathcal{S}_{\mathrm{el}} = \int_{\tau} \, \frac{m_e \dot{\mathbf{r}}_{e,\tau}^2}{2} + \frac{m_h \dot{\mathbf{r}}_{h,\tau}^2}{2} - \frac{e^2}{4\pi \varepsilon_{0}\varepsilon_{\infty} |\mathbf{r}_{e,\tau} - \mathbf{r}_{h,\tau}|}
\end{aligned}
\label{Se}
\end{equation}
where the imaginary time $\tau$ is defined over the interval 0 to $\beta \hbar$, $\beta^{-1}=k_{\textrm{B}}T$, $T$ is temperature, $k_{\textrm{B}}$ is Boltzmann's constant, and $\hbar$ is Planck's constant. The velocity and position of electron/hole are denoted $\dot{\mathbf{r}}_{e/h,\tau}$ and $\mathbf{r}_{e/h,\tau}$.
Under the assumption of a classical lattice, the contributions to the path action from MAPbI$_3$ and its interaction with the quasiparticles become $\mathcal{S}_{\mathrm{l}} = \beta \hbar \mathcal{H}_{\mathrm{l}}$ and
\begin{equation}
\begin{aligned}
\mathcal{S}_{\mathrm{int}} = \int_{\tau}^{} \, U_{e,\mathrm{l}}(\mathbf{r}_{e,\tau},\mathbf{r}^{N}) + U_{h,\mathrm{l}}(\mathbf{r}_{h,\tau},\mathbf{r}^{N})
\end{aligned}
\end{equation}
an integral over the pseudopotentials.
By discretizing the path action into a finite number of imaginary time slices, the classical counterpart of each quantum particle becomes a ring polymer consisting of beads connected by harmonic springs.\cite{habershon2013ring} 

We perform molecular dynamics (MD) simulations of two ring polymers with 1000 beads representing the electron and hole and an MAPbI$_3$ lattice with 40$\times$15$\times$15 unit cells at 300K. The large system size is necessary in order to ensure that self-interaction errors between the quasiparticles are minimized. This atomistic description allows us to capture all orders of interaction between the quasiparticles and the MAPbI$_3$ lattice, free of low temperature harmonic approximations.
The simulation details including pseudo-potentials and the force field of the lattice can be found in Supporting Information (SI \textcolor{black}{section I}).
\begin {figure}[t]
\centering\includegraphics [width=8.8cm] {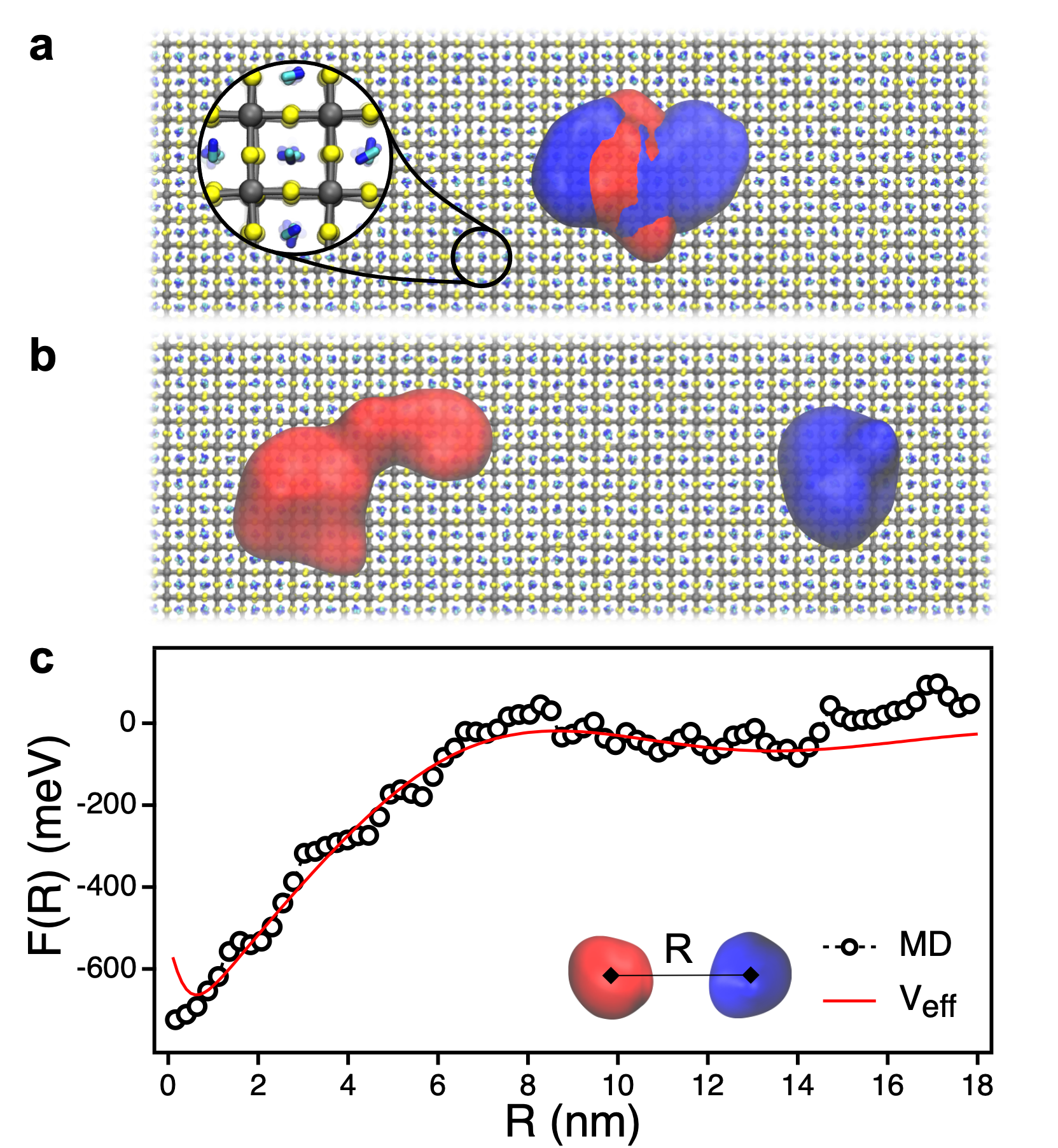}
\caption{Quasiparticle path integral molecular dynamics simulations. Representative snapshots of the simulation of electron(red) and hole(blue) with the MAPbI$_3$ lattice where electron and hole are (a) close to and (b) far from each other. Zoomed in structure in (a) represents the MAPbI$_3$ lattice where gray, yellow, and blue atoms represent Pb$^{2+}$, I$^{-}$, and MA$^+$, respectively. 
(c) Free energy between electron and hole as a function of the distance between quasiparticle centroids from molecular dynamics simulation (black circles) and an effective exciton interaction from Eq.\ref{Veff} (red solid line) with parameters as $\varepsilon^{*}=5$, $l_s = 2.47$nm, and $l_c=1.26$nm.}
\label{Fig1}
\end{figure}

To analyze the emergent exciton interaction \textcolor{black}{resulting from the collective motions} in MAPbI$_3$, we compute the free energy between electron and hole using Umbrella sampling with the Weighted Histogram Analysis Method.\cite{wham}  We compute the reversible work to move two charge centers relative to each other
\begin{equation}
\beta F(R) = -\ln \langle \delta \left (R-  |{\mathbf{r}}^{\mathrm{c}}_e - {\mathbf{r}}^{\mathrm{c}}_h| \right ) \rangle
\end{equation}
where $R$ is the distance between the electron and hole centroid ${\mathbf{r}}^{\mathrm{c}}_{e/h}$, $\delta(x)$ is Dirac's delta function, and $\langle .. \rangle$ represents an ensemble average. Simulation snapshots are shown in Figs.~\ref{Fig1}a and \ref{Fig1}b, where spatially delocalized charges extend with a radius of gyration between 1.5 - 3 nm. 
Figure~\ref{Fig1}c shows $F(R)$, which is nonmonotonic. The free energy exhibits a minima at $R=0$ reflecting the binding of the electron-hole pair into an exciton, a plateau at large $R$, and a barrier at intermediate $R\approx 8$ nm. The binding energy is large due to the neglect of polarizability in this description of the lattice.
Considering the bare Coulomb potential is a monotonic function, the repulsive interaction found in Fig.~\ref{Fig1}c at intermediate electron-hole distances must arise from the lattice. An effective electron-hole repulsion has been speculated in lead halide perovskites previously,\cite{emin2018barrier,lubin2021resolving,dana2021unusually} but had defied direct observation or theoretical validation. 


In order to understand the emergent lattice effects on the electron-hole interaction and surprising intermediate repulsion, we assume that the fluctuations of the lattice are well described by a Gaussian field. Such Gaussian field theories underpin a number of standard effective interactions including dielectric continuum theory and the Casimir effect.\cite{song1996gaussian,cox2021quadrupole,li1991fluctuation} 
A Gaussian approximation in this context is analogous to \textcolor{black}{quasi-harmonic approach \cite{karplus} where it is assumed} that while the lattice is anharmonic, it responds linearly.\cite{chandler1993gaussian,reichman2000self} We consider approximating the lattice by an effective polar displacement field,  $\mathbf{u}_{\mathbf{k},\tau}$, {\color{black}that is expected to be correlated with local bending and rocking motions of the octahedra.\cite{ferreira2020direct}} Within the Gaussian field approximation, the path action for the lattice becomes 
\begin{equation}
\begin{aligned}
\mathcal{S}_{\mathrm{l}}
&\approx \frac{1}{2}\int_{\tau}^{} \int_{\tau'}^{} \, \int_{\mathbf{k}}   \mathbf{u}_{\mathbf{k},\tau} \chi_{\mathbf{k},\tau-\tau'}^{-1} \mathbf{u}_{\mathbf{-k},\tau'} 
\end{aligned}
\end{equation}
where $\chi_{\mathbf{k},\tau-\tau'}=\langle \mathbf{u}_{\mathbf{k},\tau} \mathbf{u}_{-\mathbf{k},\tau'} \rangle $ is the susceptibility at wave vector $\mathbf{k}$ and imaginary time displacement $\tau-\tau'$. The susceptibility is determined by a phonon dispersion relationship only in the limit of zero temperature, and generally reflects the correlations within the effective polar displacement field.\cite{shih2021anharmonic} Consistent with the Coulombic pseudopotentials used in the MD simulations, we take the coupling between the charges and the lattice to be linear
\begin{equation}
\begin{aligned}
\mathcal{S}_{\mathrm{int}} \approx  \int_{\tau}^{} \, \int_{\mathbf{k}}  \,\mathbf{u}_{\mathbf{k},\tau} \, \lambda \frac{e^{i \mathbf{k} \cdot \mathbf{r}_{e,\tau}} - e^{i \mathbf{k} \cdot \mathbf{r}_{h,\tau}} }{k}
\end{aligned}
\end{equation}
and described by a Fr{\"o}hlich-like interaction,\cite{Frohlich} 
where $\lambda$ is a Fr{\"o}hlich coupling constant. 
The lattice variables can be integrated out, leaving a Gaussian approximation to the partition function, $\mathcal{Z}_{\mathrm{G}}$,
\begin{equation}
\begin{aligned}
\mathcal{Z}_{\mathrm{G}} &= \int \mathcal{D} [\mathbf{r}_e,\mathbf{r}_h, \mathbf{u}_{\mathbf{k}}] \, e^{-\mathcal{S}_{\mathrm{all}}[{\mathbf{r}}_e,{\mathbf{r}}_h,\mathbf{u}_{\mathbf{k}}] / \hbar} \\
%
&=Z_{\mathrm{l}}\int \mathcal{D} [\mathbf{r}_e,\mathbf{r}_h] \, e^{-\mathcal{S}_{\mathrm{el}}/\hbar}\, e^{-\mathcal{S}_{\mathrm{eff}} [{\mathbf{r}}_e,{\mathbf{r}}_h]/\hbar}
\end{aligned}
\end{equation}
where $Z_{\mathrm{l}}$ is the partition function for a displacement field without couplings to the charges. This integration results in an effective path action of the form (see SI for details)
\begin{equation}
\begin{aligned}
\mathcal{S}_{\mathrm{eff}} = - \sum_{i,j}
\int_{\tau}^{} \int_{\tau'}^{} 
\int_{\mathbf{k}} \Gamma_{ij} \chi_{\mathbf{k},\tau,\tau'} 
 \frac{|\lambda|^2}{2k^2}e^{i \mathbf{k} \cdot | \mathbf{r}_{i,\tau}-\mathbf{r}_{j,\tau'}|}
\end{aligned}
\label{Seff}
\end{equation}
where $i,j\in\{e,h\}$ and $\Gamma_{ij} $ takes the value of $\Gamma_{ij} = 1$ if $i = j$ and $\Gamma_{ij} = -1$ if $i \ne j$.

The susceptibility $\chi_{\mathbf{k},\tau}$ is proportional to a dielectric function evaluated in the absence of the quasiparticles. Different functional forms of its imaginary time and wavevector dependence imply different ways in which the lattice can screen the quasiparticles.    
In the classical limit\textcolor{black}{\cite{PBcl}}, $\chi_{\mathbf{k},\tau}=\chi_{\mathbf{k}} \delta( \tau)$, 
Eqs.~\ref{Se}, \ref{Seff}, and \ref{chik}, imply an effective interaction between the electron and hole, 
\begin{equation}
\hat{V}_{\mathrm{eff}}(\mathbf{k})  =- \frac{1}{k^2} \left[ \frac{e^2}{\varepsilon_0 \varepsilon_{\infty}} - \chi_{\mathbf{k}}  |\lambda |^2 \right]
\end{equation}
which is a sum of the bare interaction, here screened by $\varepsilon_{\infty}=4.5$,\cite{10.1039/c6mh00275g} and the contribution from the lattice proportional to $\chi_{\mathbf{k}}$. 
In the zero wavevector limit \textcolor{black}{this} is a constant, and if taken as $\chi_{\mathbf{k}}|\lambda|^2=e^2/\varepsilon_0 (1/\varepsilon_{\infty}-1/\varepsilon_r)$
we recover the Wannier-Mott model of a exciton. With an effective dielectric constant \textcolor{black}{$\varepsilon_{r}=6.1$},\cite{cho2019optical} this local, \emph{static} screening is manifestly insufficient to produce the repulsive interaction observed from the free energy calculations. Rather an explicit $\mathbf{k}$ dependence to $\chi_{\mathbf{k}}$ is required.

Using explicit  MD simulations of the bulk classical MAPbI$_3$ lattice, we find  $\chi_{\mathbf{k}}$ is well approximated by
\begin{equation}
\begin{aligned}
\chi_{\mathbf{k}} \approx
\frac{\chi_0}{1 - l_s^2\mathbf{k}^2 + l_s^2 l_c^2 \mathbf{k}^4} 
\end{aligned}
\label{chik}
\end{equation}
characterized by three positive real parameters, $\chi_0$, $l_s$, and $l_c$. This functional form includes a single resonant peak and is assumed isotropic on length scales greater than the lattice spacing. The resonant peak results from the negative second order coefficient and manifests the double well potential of the optical mode.\cite{physrevlett.121.086402} 
Performing the inverse Fourier transform gives an effective potential 
\begin{equation}
\begin{aligned}
&V_{\mathrm{eff}} (r) = \\
&-\frac{e^2}{4\pi \varepsilon_0 r} \left [ 
\frac{1}{\varepsilon_{\infty}} + \frac{1}{\varepsilon^*} + \frac{\gamma}{4\delta \varepsilon^*}
e^{-r\delta}\sin[r\gamma - \theta ] \right ]
\end{aligned}
\label{Veff}
\end{equation}
where $r$ is the distance between two charges.  
Details of the derivation with analytic expressions for $1/\varepsilon^*$, $\gamma^{-1}\approx \sqrt{2}\ell_c$, $\delta^{-1} \approx 2\ell_c/\sqrt{1-\ell_s/2 \ell_c}$,  and $\theta = \arctan[2\delta / \gamma]$ are shown in the SI \textcolor{black}{section II}. 
This form is plotted in Fig.~\ref{Fig1}c and provides an excellent fit at large $r$ to the free energy from the MD simulation.
We refer to this effective electron-hole interaction arising from spatially dependent screening from the MAPbI$_3$ lattice as $\textit{nonlocal}$ screening.\cite{kornyshev1981nonlocal} The theory clarifies that the non-monotonic interaction potential results from deformations generated within the lattice due to the charges. At specific characteristic distances these deformations are sufficiently unfavorable that the electron and hole are effectively repelled from each other.

{\color{black}This effective interaction in Eq.~\ref{Veff} is distinct from what has been considered previously by Pollman\textcolor{black}{n} and Buttner,\cite{pollmann1977effective} and by Gerlach and F. Luczak\cite{PBcl} in which coupling to a single dispersionless optical phonon results in a excitonic polaron that screens the bare Coulomb potential.} In their approximation, the polar displacement field is treated quantum mechanically by including a $\tau$ dependence of the susceptibility. In Pollman\textcolor{black}{n} and Buttner's work, this is taken as the bare susceptibility,
\begin{equation}
\begin{aligned}
\chi_{\mathbf{k},\tau} 
= \frac{1}{2 \omega} e^{-\omega |\tau|} 
\end{aligned}
\label{chitau}
\end{equation}
where the phonon mode is characterized by a single longitudinal optical frequency $\omega$. Plugging Eq.~\ref{chitau} into Eq.~\ref{Seff}, we obtain the effective path action
\begin{equation}
\begin{aligned}
\mathcal{S}_{\mathrm{eff}} &= -\sum_{i,j} \Gamma_{ij}
\frac{\alpha \omega_{}^2 \hbar^{3/2}}{\sqrt{8 m_{ij} \omega}}
\int_{\tau}^{} \int_{\tau'}^{} \,
\frac{e^{-\omega_{}|\tau-\tau'|}}{|\mathbf{r}_{i,\tau}-\mathbf{r}_{j,\tau'}|}
\end{aligned}
\label{Seff2}
\end{equation}
where we have written $|\lambda|^2$ in the traditional Fr{\"o}hlich form, introducing $\alpha$ as the dimensionless coupling constant. Pollman\textcolor{black}{n} and Buttner further approximate this \emph{dynamic} screening approach in order to obtain a closed form effective potential. Their potential is an exponentially screened Coulomb potential in the classical limit with a screening length given by the polaron radus.\cite{pollmann1977effective} While for certain parameters it is possible that the Pollman\textcolor{black}{n}-Buttner potential is repulsive,\cite{emin2018barrier} employing known values for $\alpha=1.72$ and $\omega=40$cm$^{-1}$ for MAPbI$_3$,\cite{10.1039/c6mh00275g} the resultant effective potential is monotonic and inconsistent with our MD result. {\color{black}Analogous approaches including sums of two or three dispersionless phonons similarly fail to describe a repulsive interaction. Treating the susceptibility variationally, Gerlach and F. Luczak\cite{PBcl} provided a more flexible description of the lattice, but the lack of a wavevector-dependence to $\chi$ still prohibits an intermediate length scale repulsion.}

\begin {figure}[t]
\centering\includegraphics [width=8.7cm] {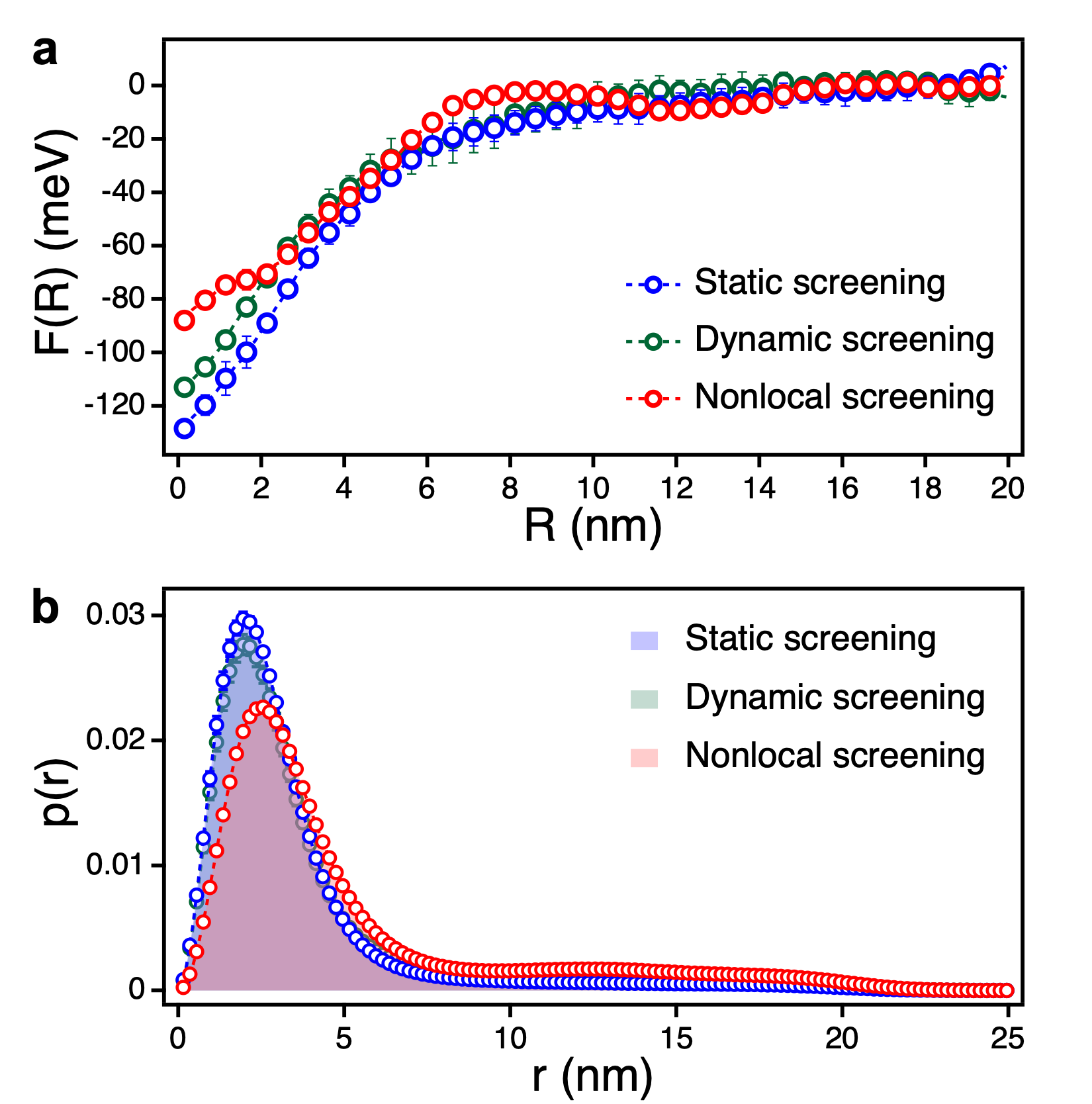}
\caption{Implications of different screening models from explicit path integral simulations with electron and hole quasiparticles. (a) Free energy as a function of the distance between quasiparticle centroids at 300K under static (blue), dynamic (green), and nonlocal (red) screening models.
(b) Radial probability distribution for electron and hole, with the same color scheme as in (a).}
\label{Fig2}
\end{figure}

To investigate the implication of a nonlocal screening on the observable properties of MAPbI$_3$, we simulated an electron and hole pair using our quasiparticle path integral approach under (i) static, (ii) dynamic, and (iii) nonlocal screenings. In each case, we employed known experimental parameters for the dielectric constants, optical frequencies, and effective masses and thus expect our results to be quantitatively accurate. To our knowledge the $\mathbf{k}$-dependent dielectric susceptibility has not been reported for MAPbI$_3$, so we parameterized the nonlocal screening interaction using our MD results. For each case, we extract the exciton binding energies and bimolecular recombination rates as both have been difficult to reconcile theoretically.\cite{herz2016charge} 

The exciton binding energy, $\Delta E_\mathrm{B}$ is definable within our path integral framework as
\begin{equation}
\Delta E_\mathrm{B} = \lim_{T\rightarrow 0} \min_{R}  \Delta F(R)
\end{equation}
and to evaluate it we computed the free energy $\Delta F(R)=F(R)-F(\infty)$ at a variety of temperatures  ranging from 200K to 400K and extrapolate its value to 0K (SI \textcolor{black}{Fig.S2}). Representative free energies at $T=300$K are shown in Fig.~\ref{Fig2}a. As anticipated from the theory, we find that both dynamic and nonlocal screening reduce the effective attraction between electron and hole but only the nonlocal screening results in a barrier to recombination. 

The extrapolated binding energies are summarized in Table \ref{table1}. Within the static screening approach, the exciton is hydrogenic, and the binding energy is given by $\Delta E_\mathrm{B}^{s} = \mu \textcolor{black}{e}^4/2 (4\pi \epsilon_0 \epsilon_r)^2 \hbar^2$ where $\mu$ is a reduced mass of the electron and hole. The large decrease in binding energy under dynamic screening reflects the polaronic effect. Since the experimentally derived value of $\alpha$ is relatively small,\cite{10.1039/c6mh00275g} we find the change to the binding energy is well approximated by first order perturbation theory, yielding the known Fr{\"o}hlich result, $\Delta E_\mathrm{B}^{d}=\Delta E_\mathrm{B}^{s} -2\alpha \hbar \omega$.\cite{ptemass} This reduction in the binding energy is consistent with recent Bethe-Salpeter calculations with perturbative electron-phonon interactions,\cite{neatonprl} but higher than experimental estimates.\cite{galkowski2016determination}  The reduction in the binding energy from the nonlocal screening is 12 meV, which is close to a prediction assuming hydrogenic 1s orbits, 17 meV. 

\textcolor{black}{In the low temperature limit, the classical lattice approximation employed to construct the nonlocal screening model is no longer valid. In this limit, quantization of the phonons can lead to hybridization and polaron formation. To estimate the quantum mechanical effect of phonons in this model, we have adopted a hybrid approach where we have added a single optical phonon as done in the dynamical approximation, to the effective potential description deduced form the classical lattice simulations. The dynamical mode is treated analogously as Eq. \ref{Seff2}, while the effective potential is assumed to be constant at low temperatures and reflective of dynamic disorder. Treating both of these effects yields a binding energy of 20.8 meV in very good agreement with experiment.\cite{galkowski2016determination, 10.1038ncomms4586} }

The bimolecular recombination rate, $k_r$ is defined as the rate of change of the concentration of free charges, 
\begin{equation}
\frac{d\rho_e}{dt} = -k_r \rho_e \rho_h
\end{equation}
through the reaction $e^{-}+h^{+}\rightarrow \hbar \nu$, where $\rho_{e/h}$ is the concentration of free electrons/holes. At typical working excitation densities for MAPbI$_3$ based photovoltaics, radiative recombination is the limiting factor determining the charge carrier lifetime.\cite{herz2016charge} We can evaluate $k_r$ using Fermi's golden rule for spontaneous emission, with an effective mass approximation.\cite{ratebook1, ratebook2, rate1.4804183, rate1.1623330} Within our quasiparticle path integral framework, the rate is given by a constant times a ratio of path partition functions,\cite{physrevb.73.165305}
\begin{equation}
k_{\mathrm{r}} = \frac{e^2 \sqrt{\varepsilon_{\infty}} E_{\mathrm{gap}}^2}{2 \pi \varepsilon_0 \hbar^2 c^3 \mu } \frac{{\mathcal{Z}}_{\mathrm{c}}}{\mathcal{Z}}
\end{equation}
where $E_{\mathrm{gap}}=1.64e$V is the band gap energy for MAPbI$_3$,  and $c$ is the speed of light. The subscript $\mathrm{c}$ on $\mathcal{Z}_{\mathrm{c}}$ stands for combined path integral in which the two separate imaginary time paths are placed together to form a single, radiating path by linking same imaginary time slices. 
The ratio of path partition functions can be evaluated as,
\begin{equation}
\frac{\mathcal{Z}_{\mathrm{c}}}{\mathcal{Z}_\mathrm{}} = 4 \pi  
\int dR\, R^2 \left \langle e^{\Delta S/\hbar} \right \rangle_R  e^{-\beta \Delta F(R)}
\end{equation}
where we replace the ratio of partition functions by an exponential average at fixed $R$ of the difference in path action $\Delta S = S - S_{\mathrm{c}}$ (SI \textcolor{black}{Eq.S13}).

\renewcommand{\arraystretch}{1.5}
\begin{table}[t]
\centering
\begin{tabular}{ |P{2.25cm} |P{1.8cm} | P{1.8cm} |P{1.8cm}| }
\hline 
Screenings & $ \textrm{Static} $  &  $\textrm{Dynamic}$ & $\textrm{Nonlocal}$ \\
\hline
$\Delta \rm{E}_B \, (\rm{meV})$ & \textcolor{black}{50.4} & \textcolor{black}{36.9} & \textcolor{black}{38.1} \\
\hline
$\tau_{\mathrm{r}} \, (\rm{ns})$ & 13.5 & 35.5 & 78.1 \\
\hline 
\end{tabular}
\caption{Exciton binding energy $\Delta E_{\mathrm{B}}$, and carrier lifetime $\tau_{\mathrm{r}}$ estimated using different screening models.}
\label{table1}
\end{table}

The change in action is a reporter on the overlap between the electron and hole wavefunctions. The electron and hole radial probability distribution is described in Fig.~\ref{Fig2}b, illustrating that the nonlocal screening effect increases the average distance between electron and hole. This is distinct from the effect of the dynamic screening, which consistent with the small value of $\alpha$, leaves the electronic distribution largely unaltered from the simple Wannier, static screening model. The decrease in electron hole overlap results in a nearly order of magnitude decrease in $k_\mathrm{r}$ using the nonlocal screening theory relative to the static screening theory. For the nonlocal screening theory, we find $k_\mathrm{r}=1.3 \times 10^{-10}$cm$^3/$s, in excellent agreement with photoluminescence lifetime measurements.\cite{herz2016charge}

Assuming the only loss mechanism is due to bimolecular recombination, the lifetime of an electron hole pair is computable from $\tau_\mathrm{r} = 1/k_\mathrm{r} \rho_e$. Summarized in Table 1 for $\rho_e=10^{17}/$cm$^3$ are lifetime estimates using the different screening models. Both standard polaronic effects incorporated into the dynamic screening model as well as the nonlocal screening model increase the lifetimes of free charge carriers, however the contribution of a nonlocal screening obtained from the MAPbI$_3$ lattice is much more significant. The details on the bimolecular recombination rate calculations are in the SI \textcolor{black}{section III}.

By comparing these different simplified models that account for charge-lattice interactions, we find that the ability of the MAPbI$_3$ lattice to nonlocally screen quasiparticles is sufficient to explain the particularly low exciton binding energy and recombination rate.  Only this screening kernel in our unified Gaussian field theory formalism can suppress the electron-hole overlap enough to explain the anomalously long free carrier lifetime with weak lattice coupling strength. The particular nonlocal screening adopted here was deduced directly from explicit atomistic molecular dynamics simulations using a quasiparticle path integral framework. This framework is uniquely able to study the thermodynamics of this quasiparticle-lattice system at finite temperature. 

The adoption of a spatially dependent screening is consistent with a growing literature pointing to the importance of dynamic disorder in lead halide perovskites.\cite{leguy2016dynamic,schilcher2021significance, quan2021vibrational} {\color{black}As it is the wave-vector dependent dielectric susceptibility of the bulk ground state lattice that enters into the theory presented above, experimental measurements of such properties could afford a means of assessing  potential materials with similarly long radiative lifetimes.} 
Further, the barrier to bringing electrons and holes together we have discovered here undoubtedly has implications apart from offering an explanation of the particular high power conversion efficiencies of MAPbI$_3$. For example, this repulsion may help explain observations of anti-binding of biexcitons.\cite{lubin2021resolving, dana2021unusually} The identification of a repulsive electron-hole interaction generated from the soft, polar modes of the perovskite lattice offers a key new design principle for photovoltaic materials. Searching for other systems that admit this type of interaction represents a promising new direction for  materials discovery.


\textcolor{black}{The Supporting material is available free of charge via the internet at http://pubs.acs.org, which includes the details of simulations, derivation of effective electron-hole interaction, and the bimolecular recombination rate with path integral framework.}

\vspace{1mm}
{\bf Acknowledgments.} This work was supported by the U.S. Department of Energy, Office of Science, Office of Basic Energy Sciences, Materials
Sciences and Engineering Division under Contract No. DE-AC02-05-CH11231 within the Physical Chemistry of Inorganic Nanostructures Program (No. KC3103). This research used resources of the National Energy Research Scientific Computing Center (NERSC), a U.S. Department of Energy Office of Science User Facility. Y. P. also acknowledges Kwanjeong Educational Foundation. D. T. L. acknowledges the Alfred P. Sloan Foundation.  

\vspace{1mm}
{\bf References}

\bibliography{main}

\end{document}


\title{Supporting Information to ``Nonlocal screening dictates the radiative lifetimes of excitations
in the lead halide perovskites"}

\author{Yoonjae Park}
 \affiliation{Department of Chemistry, University of California, Berkeley}

\author{Amael Obliger}
 \thanks{Current Address: ISM, Univ. Bordeaux, CNRS, Talence, France}
 \affiliation{Department of Chemistry, University of California, Berkeley}
 
\author{David T. Limmer}
 \email{dlimmer@berkeley.edu}
 \affiliation{Department of Chemistry, University of California, Berkeley}
\affiliation{Materials Science Division, Lawrence Berkeley National Laboratory}
\affiliation{Chemical Science Division, Lawrence Berkeley National Laboratory}
\affiliation{Kavli Energy NanoScience Institute, Berkeley, California, Berkeley}

\date{\today}

\maketitle

\section{Simulation details}

\subsection{$\textrm{MAPbI}_{\textbf{3}}$ lattice with electron and hole ring polymers}

We consider MAPbI$_3$ lattice of 40$\times$15$\times$15 unit cells with electron and hole ring polymers. For the atomistic model of MAPbI$_3$ lattice, we adopt empirical force field from Ref~\cite{mattoni}. Interactions between inorganic atoms (Pb, I) are described by Buckingham-Coulomb (BC) potential
\begin{equation}
U_{\mathrm{BC}} = \sum_{i,j} A_{ij} e^{-r_{ij} / \rho_{ij}} - \frac{c_{ij}}{r_{ij}} 
+ \frac{q_i q_j}{4 \pi \varepsilon_0 r_{ij}}
\label{buck}
\end{equation}
where $r_{ij}$ is the distance between $i$ and $j$ atoms, $\varepsilon_0$ is the vacuum electric permittivity, $q_i$ is the charge of atom $i$, and $A_{ij}$, $c_{ij}$, and $\rho_{ij}$ are parameters for BC potential.
Other pairwise interactions are described by the sum of BC potential and Lennard-Jones (LJ) potential
\begin{equation}
U_{\text{LJ}} = \sum_{i,j} 4 \varepsilon_{ij} \Big[ \Big( \frac{\sigma_{ij}}{r_{ij}} \Big)^{12} - \Big( \frac{\sigma_{ij}}{r_{ij}} \Big)^6 \Big]
\label{LJ}
\end{equation}
with LJ parameter $\varepsilon_{ij}$ and $\sigma_{ij}$. Types and parameters for pairwise interactions can be found in Table~\ref{buckff} and Table~\ref{LJff}. 
For electron and hole ring polymers, we used a quasiparticle path integral approach whose Hamiltonian obtained by discretizing the path action 
is given by 
\begin{equation}
\begin{aligned}
\mathcal{H}_{\mathrm{RP}} = \sum_{j=1}^{n} \frac{m_en}{2\beta^2\hbar^2}(\mathbf{r}_{e,j+1}-\mathbf{r}_{e,j})^2 + \sum_{j=1}^{n} \frac{m_hn}{2\beta^2 \hbar^2}(\mathbf{r}_{h,j+1}-\mathbf{r}_{h,j})^2
- \sum_{j=1}^n \frac{e^2 }{4\pi \varepsilon_0 n |\mathbf{r}_{e,j} - \mathbf{r}_{h,j}|}
\end{aligned}
\label{HRP}
\end{equation}
where $\mathbf{r}_{e/h,j}$ is the position of the $j^{\mathrm{th}}$ bead in electron/hole ring polymer, $n=1000$ is the number of beads used in each ring polymer, $\beta=(k_{\textrm{B}}T)^{-1}$ is an inverse temperature, and $m_e/m_0 = m_h/m_0 = 0.2$\cite{jpcl9b02491} are the band masses of electron and hole with the bare electron mass $m_0$. 
Pseudopotentials are given by truncated Coulomb potentials where $r^{-1}$ is replaced by $(\alpha_{C} + r^2 )^{-1/2}$. The parameter $\alpha_{C}$ for electron-lattice, hole-lattice, and electron-hole interactions are chosen to recover the corresponding ionization energy, electron affinity, and the band gap of MAPbI$_3$ perovskite, summarized in Table~\ref{alphac} \cite{pseudo1, pseudo2}. 

Simulations are run in NVT ensemble with Langevin thermostat to control the temperature to 300K using timestep 0.5fs.
%
%
In the free energy calculations using Umbrella sampling\cite{usamp1},  
we used 86 windows where the harmonic potentials $V(R) = 0.5k_{\mathrm{sp}}(R-R_{\mathrm{eq}})^2$ are added to the distance between two centroids of ring polymers $R_{}$ with the spring constant $k_{\mathrm{sp}}$ ranging from $0.15$ to $0.4 \,\rm{kcal/mol/}\rm{\AA}^2$.
Prior to adding the ring polymers to the lattice, they are equilibrated for 500ps in NVT ensemble with Langevin thermostat at each window as is done in the staging algorithm\textcolor{black}{\cite{sprik1985staging}}. We combine the equilibrated configuration of ring polymers with the lattice, run 200ps for equilibration purpose, and store trajectories in every 50fs for 700ps. 
%



\subsection{Electron and hole ring polymers under three types of screenings}

Simulations with isolated electron and hole ring polymers are performed under each type of effective screening. For a $\textit{static}$ screening, the discretized Hamiltonian is given by Eq.\ref{HRP} with Coulomb interaction screened by dielectric constant $\varepsilon_r = 6.1$ \cite{dielec}. 
For the $\textit{dynamic}$ screening, Hamiltonian consists of two pieces, $\mathcal{H_{\mathrm{dyn}}} = \mathcal{H}_{\mathrm{RP}} + \mathcal{H}_{\mathrm{eff}}$. The second term is given by Eq.~15 with the factor of discretization $n$,
\begin{equation}
\mathcal{H}_{\mathrm{eff}} = -\sum_{i,j \in \{ e,h \}} \Gamma_{ij}
\frac{\alpha \beta \hbar^2 \omega^2}{n^2} \sqrt{\frac{\hbar }{8m_{ij} \omega}} 
\sum_{t=1}^n \sum_{s=1}^n 
\frac{e^{- \frac{\beta \hbar \omega}{n} |t-s|}}{\big| \mathbf{r}_{i,t} - \mathbf{r}_{j,s} \big| } 
\label{1rpH}
\end{equation}
where $\mathbf{r}_{e/h,j}$ represents the position of the $j^{\mathrm{th}}$ bead in electron/hole ring polymer, $\Gamma_{ij} = 1$ if $i=j$ and $\Gamma_{ij}=-1$ if $i\ne j$,  $\alpha = 1.72$ \cite{10.1039/c6mh00275g} is Fröhlich coupling constant and $\omega=\rm{40cm}^{-1}$\cite{10.1039/c6mh00275g} is optical frequency of MAPbI$_3$ perovskite with $m_{eh} = \mu$. 
For the $\textit{nonlocal}$ screening, the last term in Eq.\ref{HRP} is replaced by the effective exciton interaction from empirical force field calculation shown in Fig.1c (black symbols), divided by the effective dielectric constant \textcolor{black}{$\varepsilon_r$} to account for the lack of explicit polarizability in the force field.

Simulations are run in NVT ensemble and Langevin thermostat with pseudopotential.
For Umbrella sampling parameters, 65 windows are used with the spring constant $k_{\mathrm{sp}}=0.2 \rm{kcal/mol/}\rm{\AA}^2$ and equilibrium distance $R_{\mathrm{eq}}$ starts from $3\rm{\AA}$ and is increased by $3\rm{\AA}$ at each window.
In the first window, ring polymers are first equilibrated for 1ns with 1fs timestep and then trajectories are stored in every 500ps for 10ns. 
The last configuration of the previous window is taken to be the starting point of the next window. 
%
%
Simulations are performed using the LAMMPS package\cite{lammps}.

\section{Derivation of an effective electron-hole interaction}

Starting from the Gaussian approximated path action given by $\mathcal{S}_{\mathrm{l}} + \mathcal{S}_{\mathrm{int}}$ from Eq.7 and Eq.8, its Fourier transform becomes
\begin{equation}
\begin{aligned}
\hat {\mathcal{S}}_{\mathrm{l}} + \hat{\mathcal{S}}_{\mathrm{int}}
= \frac{1}{2} \int_{\mathbf{k}}\int_{\omega}^{} \, \chi_{\mathbf{k},\omega}^{-1} |\mathbf{u}_{\mathbf{k},\omega}|^2 
 + \int_{\mathbf{k}} \int_{\tau=0}^{\beta \hbar}
\int_{\omega} \mathbf{u}_{\mathbf{k},\omega}  \,e^{-i\omega \tau}\, \lambda \frac{e^{i \mathbf{k} \cdot \mathbf{r}_{e,\tau}} - e^{i \mathbf{k} \cdot \mathbf{r}_{h,\tau}} }{k}
\end{aligned}
\end{equation}
where a Fröhlich-type coupling constant $\lambda$ is defined as 
\begin{equation}
\begin{aligned}
\lambda = -i \hbar \omega \left( \frac{4 \pi \alpha }{V} \right)^{\frac{1}{2}} \left( \frac{\hbar}{2m\omega} \right)^{\frac{1}{4}} \left( \frac{2\omega }{\hbar}\right)^{\frac{1}{2}}
\end{aligned}
\label{Fint}
\end{equation}
with dimensionless coupling constant $\alpha$\cite{Anonymous:8b6}.
Using the fact that the Gaussian integral of the form of $e^{-x^2/2\sigma^2 + ax}$ with respect to $x$ produces the term of $e^{\sigma^2 |a|^2/2}$ gives the effective path action defined in Eq.9 as
\begin{equation}
\begin{aligned}
\mathcal{S}_{\mathrm{eff}} 
&= -\frac{1}{2} \int_{\mathbf{k}} \int_{\omega} \chi_{\mathbf{k},\omega}
\left| \int_{\tau=0}^{\beta \hbar} e^{-i\omega \tau}\, \lambda \frac{e^{i \mathbf{k} \cdot \mathbf{r}_{e,\tau}} - e^{i \mathbf{k} \cdot \mathbf{r}_{h,\tau}} }{k} \right|^2 \\[3pt]
&= -\frac{1}{2} \int_{\tau=0}^{\beta \hbar} \int_{\tau'=0}^{\beta \hbar} \int_{\mathbf{k}} \int_{\omega} 
\chi_{\mathbf{k},\omega} e^{-i\omega (\tau-\tau')}
\frac{| \lambda |^2}{k^2}
\left|  e^{i \mathbf{k} \cdot \mathbf{r}_{e,\tau}} - e^{i \mathbf{k} \cdot \mathbf{r}_{h,\tau}} \right|^2 \\[3pt]
&= -\frac{1}{2} \int_{\tau=0}^{\beta \hbar} \int_{\tau'=0}^{\beta \hbar} \int_{\mathbf{k}} \chi_{\mathbf{k},\tau - \tau'} 
\frac{| \lambda |^2}{k^2}
\left|  e^{i \mathbf{k} \cdot \mathbf{r}_{e,\tau}} - e^{i \mathbf{k} \cdot \mathbf{r}_{h,\tau}} \right|^2
\end{aligned}
\end{equation}
where inverse Fourier representation of $\chi_{\mathbf{k},\omega}$ is used in the third line, which can be reduced to Eq.10 in the compact form. 
%

\begin {figure*}[h]
\centering\includegraphics [height=5.3cm] {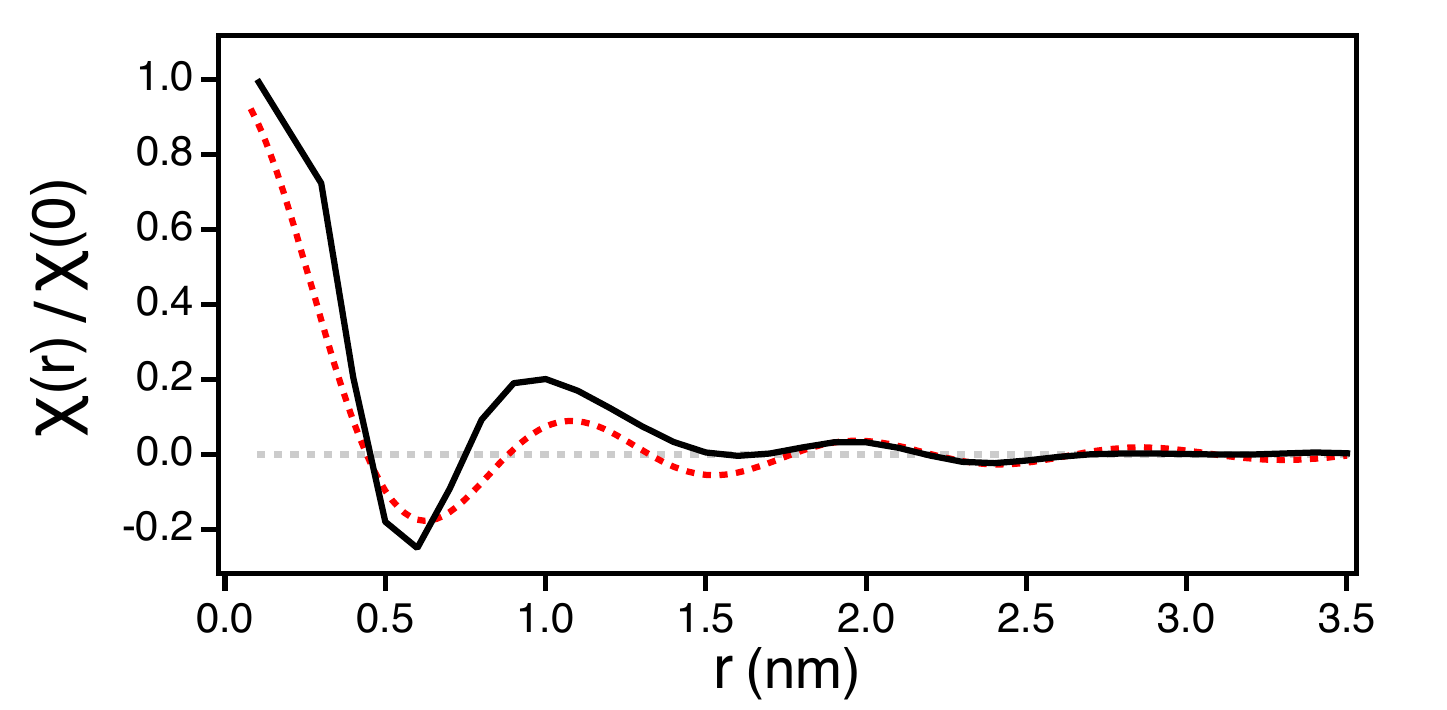}
\caption{Susceptibility $\chi (r)$ from explicit MD simulation (black solid line) and from Eq.\ref{chir} (red dotted line). Gray dotted line is the guideline for 0.}
\vspace{7mm}
\label{chi}
\end{figure*}

In the classical limit, taking $\chi_{\mathbf{k},\tau}=\chi_{\mathbf{k}} \delta( \tau)$ removes the $\tau$ dependence in the effective path action and gives an effective electron-hole interaction as Eq.11. 
We perform explicit MD simulations of MAPbI$_3$ lattice without electron and hole quasiparticles and compare $\chi (r)$ in real space where an analytic expression can be derived from the inverse Fourier transform using residue theorem with $\gamma$ and $\delta$ defined in Eq.\ref{para}.
\begin{equation}
\begin{aligned}
\chi (r) = \frac{\chi_0}{4\pi r} \frac{(\gamma^2 + \delta^2)^2}{4\gamma \delta} e^{-r\delta } \sin[r\gamma]  
\end{aligned}
\label{chir}
\end{equation}
Fig.\ref{chi} shows $\chi(r)$ from MD simulation (black solid line) and from Eq.\ref{chir} (red dotted line) where the reasonable agreement shows the validity of the functional form of $\chi_{\mathbf{k}}$ given by Eq.12.
Using Eq.11 and 12, performing inverse Fourier transform with residue theorem gives 
\begin{equation}
\begin{aligned}
V_{\mathrm{eff}} (r) =
 -\frac{e^2}{4\pi \varepsilon_0 r} \left [ \frac{1}{\varepsilon_{\infty}}  + \frac{1}{\varepsilon^*} - \frac{e^{-r\delta}}{2 \varepsilon^*} \left( \cos[r\gamma] -
\frac{\gamma^2 - \delta^2}{2 \gamma \delta}\sin[r\gamma]  \right) \right ]
\end{aligned}
\end{equation}
where parameters, $\gamma$ and $\delta$ are defined as follows.
\begin{equation}
\begin{aligned}
\frac{1}{\varepsilon^*} = \frac{\chi_0 |\lambda|^2 \varepsilon_0}{e^2}  \quad , \quad
\gamma = \sqrt{\frac{1}{2l_s l_c} + \frac{1}{4l_c^2}} \quad , \quad 
\delta = \sqrt{\frac{1}{2l_s l_c} - \frac{1}{4l_c^2}} \quad 
\end{aligned}
\label{para}
\end{equation}
Since $\gamma \gg \delta$, the relation, $a \sin \theta - b \cos \theta = \sqrt{a^2 + b^2 } \sin [\theta - \arctan [b/a]]$, simplifies the effective electron-hole interaction as given in Eq.13.

\section{Bimolecular recombination rate with path integral}

The bimolecular recombination rate derived from Fermi's golden rule under effective mass approximation can be computed from the ratio of partition functions\cite{physrevb.73.165305, ratebook1, ratebook2, rate1.4804183, rate1.1623330}
\begin{equation}
\begin{aligned}
k_{\mathrm{r}} = \frac{e^2 \sqrt{\varepsilon_{\infty}} E_{\mathrm{gap}}^2}{2 \pi \varepsilon_0 \hbar^2 c^3 \mu } \frac{{\mathcal{Z}}_{\mathrm{c}}}{\mathcal{Z}_{\mathrm{}}}
\end{aligned}
\label{brate}
\end{equation}
where $e$ is the charge of an electron, $\varepsilon_0$ is vacuum electric permittivity, $\varepsilon_{\infty}$ is the dielectric constant related to the index of refraction, $\hbar$ is Planck's constant, $c$ is the speed of light, $\mu$ is the reduced mass of electron and hole, and $E_{gap}=1.64\textrm{eV}$ is the band gap energy for MAPbI$_3$ perovskite. 
In the path integral framework, $\mathcal{Z}$ indicates the path for two separate (electron and hole) quasiparticles and $\mathcal{Z}_c$ stands for the combined path where electron and hole paths are combined forming a radiating path by linking the same imaginary time slices. 
$\mathcal{Z}_c$ can be rewritten in terms of $\mathcal{Z}$ and the difference in path action $\Delta S = S - S_c$, 
\begin{equation}
\begin{aligned}
\mathcal{Z}_{c} &= \int \mathcal{D}[\mathbf{r}_e, \mathbf{r}_h] \, e^{-S_{c}+S-S} \, \frac{\mathcal{Z}}{\mathcal{Z}} 
= \mathcal{Z} \int \mathcal{D}[\mathbf{r}_e, \mathbf{r}_h] \, \frac{e^{-S}}{\mathcal{Z} } \, e^{-S_{c} + S_{}} 
= \mathcal{Z} \, \langle e^{\Delta S} \rangle_{}
\end{aligned}
\end{equation}
where $S$ and $S_c$ are the path actions of two separate and combined paths. The change in path action $\Delta S$ is give by
$$
\Delta S =  \int_{\tau=0}^{\beta \hbar} \frac{m_e \dot{\mathbf{r}}_{e,\tau}^2}{2} + \frac{m_h \dot{\mathbf{r}}_{h,\tau}^2}{2} - \frac{\mu \dot{\mathbf{r}}_{\mathrm{c},\tau}^2}{2}
$$
where $\mathbf{r}_{\mathrm{c}}$ is a \textcolor{black}{combined} path integral where the two paths are shown in Figs.~\ref{rate}b and c. 
 With the conditional probability representation, 
%
the ratio of partition functions becomes
\begin{equation}
\begin{aligned}
\frac{\mathcal{Z}_c}{\mathcal{Z}} 
= \langle e^{\Delta S} \rangle_{}
&= \int d\Delta S \ P_{}(\Delta S) \ e^{\Delta S}
= \int d\Delta S \left[ 4\pi \int d{R} \,R^2 P(\Delta S | {R}) \, P({R}) \right]  e^{\Delta S} \\[2pt]
&= 4\pi \int dR \, R^2 \int d\Delta S \, P(\Delta S | {R}) \, e^{\Delta S} P({R})
= 4\pi \int dR \, R^2 \langle e^{\Delta S} \rangle _R \, P(R)
\end{aligned}
\label{expdS}
\end{equation}
where $\langle e^{\Delta S} \rangle _R$ indicates the value of $\langle e^{\Delta S} \rangle$ given R and $P(R)$ can be computed from the free energy.
Plugging Eq.\ref{expdS} into Eq.\ref{brate} results in Eq.19. 
Shown in Fig.\ref{rate}a is the ratio of partition function with various number of beads. Since the error of Trotter factorization in path integral approach scales by $n^{-2}$, we compute the carrier lifetime using the value extrapolated to $n \rightarrow \infty$.

\vspace{5mm}
\bibliography{si}


\newpage
\begin{table}[h]
\centering
\begin{tabular}{ |P{1.0cm} |P{1.2cm}| P{3.3cm}| P{2.8cm}| P{3cm}| }
\hline
$i$  & $j$ &  $A_{ij}$ (kcal/mol) &  $\rho_{ij} $ ($\mathrm{\AA}$)  &  $c_{ij}$ (kcal/mol/$\mathrm{\AA}^6$) \\
\hline
 Pb &  Pb & 70359906.629702 & 0.131258 & 0.0000 \\
 Pb &  I & 103496.133010 & 0.321737 & 0.0000 \\
 Pb &  C/N & 32690390.937995 & 0.150947 & 0.0000 \\ 
 I &  I & 24274.905590 & 0.482217 & 696.949542 \\
 I &  C/N & 112936.714213 & 0.342426 & 0.0000 \\
\hline
\end{tabular}
\caption{Parameters for Buckingham potential (Eq. \ref{buck}).}
\vspace{5mm}
\label{buckff}
\end{table}

\begin{table}[h]
\centering
\begin{tabular}{ |P{1.0cm} |P{1.0cm}| P{2.7cm}| P{2.0cm}| }
\hline
$i$  & $j$ & $\varepsilon_{ij}$ (kcal/mol) & $\sigma_{ij} $ ($\mathrm{\AA}$) \\
\hline
Pb & H$_n$ & 0.0140 & 2.26454  \\
Pb & H & 0.0140 & 2.70999   \\
I & H$_n$ & 0.0574 & 2.750  \\
I & H & 0.0574 & 3.100  \\
C & C & 0.1094 & 3.39970   \\
C & N & 0.1364 & 3.32480   \\
C & H$_n$ & 0.0414 & 2.23440   \\
C & H & 0.0414 & 2.67980   \\
N & N & 0.1700 & 3.250   \\
N & H$_n$ & 0.0517 & 2.15950   \\
N & H & 0.0517 & 1.0690   \\
H$_n$ & H$_n$ & 0.0157 & 1.06910   \\
H$_n$ & H & 0.0157 & 1.51450   \\
H & H & 0.0157 & 1.960   \\
\hline
\end{tabular}
\caption{Parameters for Lennard-Jones potential (Eq. \ref{LJ}). The hydrogen atoms in ammonium NH$_3^+$ group are indicated by H$_n$.}
\vspace{5mm}
\label{LJff}
\end{table}

\begin{table}[h]
\centering
\begin{tabular}{ |P{1.0cm} |P{1.7cm}| P{1.8cm}| P{1.8cm}| P{1.8cm}| }
\hline
$\alpha_C$  & $ \textrm{Pb} $  &  $\textrm{I}$ & $b_e$ & $b_h$ \\
\hline
$b_e$ & 3.67 & 0.005 & - & 2.072 \\ 
$b_h$ & 0.005 & 22.152 & 2.072 & - \\ 
\hline
\end{tabular}
\caption{Pseudopotential parameter $\alpha_C$ where $b_{e(h)}$ indicates the bead in the electron(hole) ring polymer.}
\label{alphac}
\end{table}

\newpage

\newpage
\begin {figure*}[h]
\centering\includegraphics [height=7.5cm] {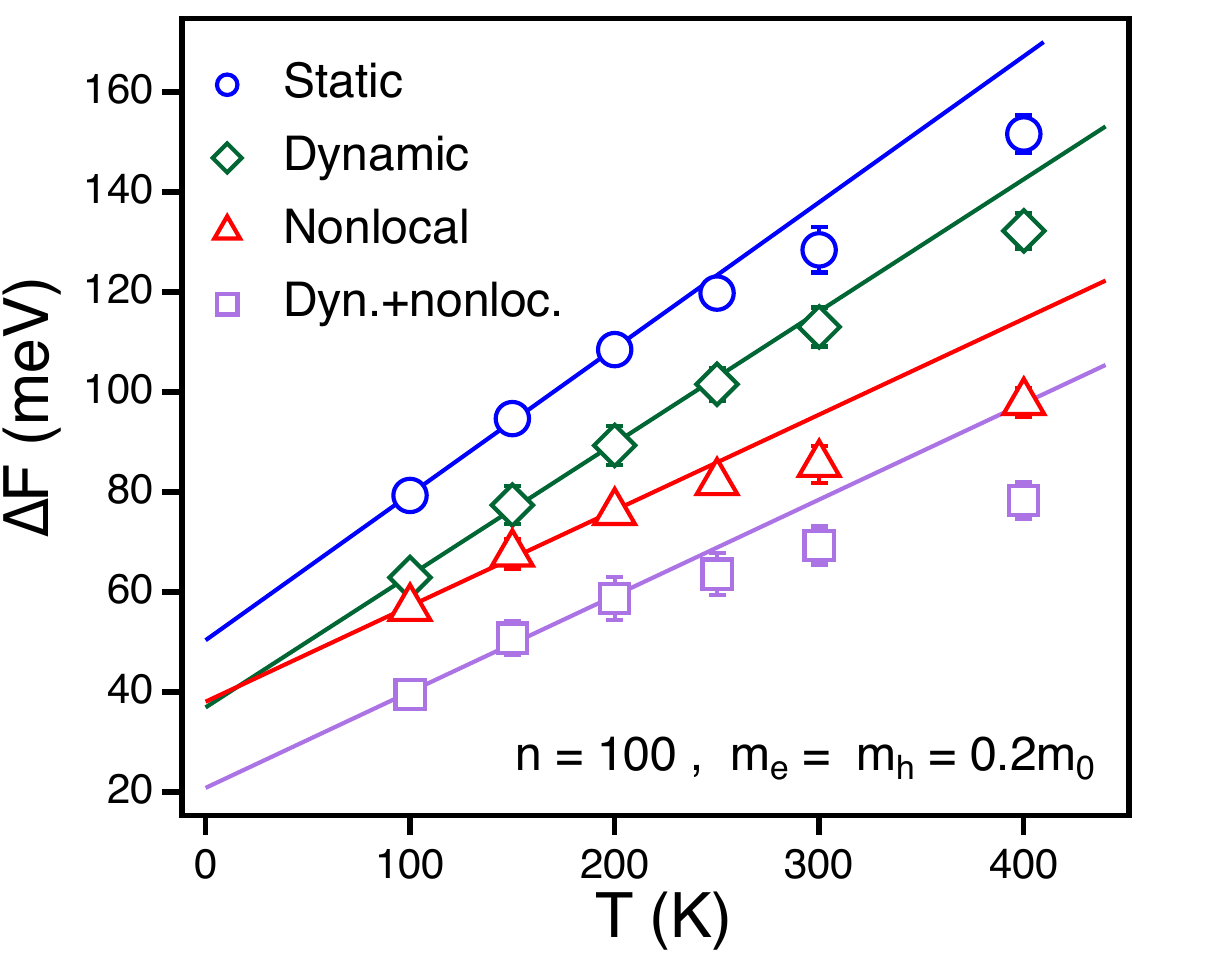}
\caption{The difference in free energy $\Delta F$ as a function of temperature under static screening (circles), dynamic screening (diamonds), nonlocal screening (triangles), and nonlocal screening combined with polaronic effect (squares). \textcolor{black}{Solid} lines are corresponding fitted lines.}
\label{bindE}
\end{figure*}

\newpage
\begin {figure*}[h]
\centering\includegraphics [height=6.5cm] {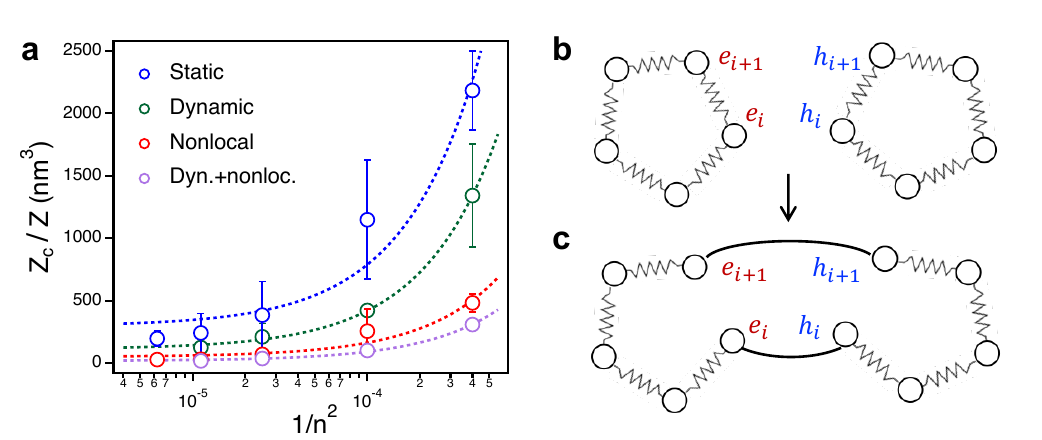}
\caption{(a) Ratio of partition function with various number of beads. Dotted lines are linear fittings under each type of screening.  
Schematic picture of (b) two separate paths $S$ and (c) combined path $S_c$. }
\label{rate}
\end{figure*}



